\documentclass[conference]{IEEEtran}
\IEEEoverridecommandlockouts

\usepackage{cite}
\usepackage{amsmath,amssymb,amsfonts}
\usepackage{algorithmic}
\usepackage{graphicx}
\usepackage{caption}
\usepackage{subcaption}
\usepackage{textcomp}
\usepackage{xcolor}
\usepackage{multirow}
\usepackage{enumitem}
\usepackage{algorithm}
\usepackage{comment}
\usepackage{array}
\usepackage{mathtools}

\def\BibTeX{{\rm B\kern-.05em{\sc i\kern-.025em b}\kern-.08em
    T\kern-.1667em\lower.7ex\hbox{E}\kern-.125emX}}

\begin{document}

\title{Nonlinear Equalization for TDMR Channels \\ Using Neural Networks
\thanks{\textsuperscript{$*$}Work done while the author was with Marvell Semiconductor Inc.}
}

\author{%
\IEEEauthorblockN{1\textsuperscript{st} Jinlu Shen}
\IEEEauthorblockA{
\textit{Washington State University}\textsuperscript{$*$}, \\ Pullman, WA 99164 USA\\
jinlu.shen@wsu.edu}
\and
\IEEEauthorblockN{2\textsuperscript{nd} Nitin Nangare,}
\IEEEauthorblockA{\textit{Marvell Semiconductor Inc.}, \\ Santa Clara, CA 95054 USA\\
nitinn@marvell.com}%
}

\maketitle

\begin{abstract}
This paper presents new structure and adaptation criterion for equalization of two-dimensional magnetic recording channels, as opposed to typical linear equalizer with minimum mean square error (MMSE) as adaptation criterion. To compensate for the nonlinear channel noise, we propose a neural network based nonlinear equalizer and show it outperforms linear equalizer under the same criterion. To achieve minimum bit error rate (BER) at the detector output, we propose to adapt the equalizer with cross entropy between the true probability of the bit and detector's estimate of it. We show minimizing the cross entropy 
enables maximum likelihood adaptation,
and results in lower detector BER than the MSE criterion. Several variations of nonlinear equalizer structures with cross entropy criterion are investigated. Compared to linear MMSE equalizer, the proposed scheme can provide up to 22.76\% detector BER reduction with only 6$\times$ increase in complexity.
\end{abstract}


\section{Introduction}
In digital magnetic recording, partial response maximum likelihood (PRML) signaling techniques are typically adopted in the read channel \cite{prml, tdmr}, in which a linear equalizer (LE) with partial response (PR) signaling is followed by a maximum likelihood (ML) sequence detector, such as the Viterbi algorithm \cite{viterbi}. The equalizer shapes the channel response to a shorter duration, thereby limiting the complexity of the detector. Such a scheme is used to combat extensive inter-symbol interference (ISI), especially at higher recording densities. Currently, linear minimum mean square error (MMSE) equalizers with finite-impulse-response (FIR) are widely used, which minimize the mean square error (MSE) between the equalized output and desired response.
In high density recording channels, a read channel technology called two-dimensional magnetic recording (TDMR) \cite{Roger} can be employed, in which two or more readers are placed on top of one or more closely packed tracks, thereby enabling 2D equalization of inter-track interference (ITI). However, linear MMSE equalization is, in general, suboptimal, in terms of both equalizer structure and adaptation criterion. Firstly, magnetic recording channels inherently suffer from nonlinearity, including transition shift and broadening, partial erasure, pattern-dependent noise, ISI, ITI, and asymmetry \cite{mr_overview}. Suitable for ISI channels with additive white Gaussian noise (AWGN), linear equalization is unable to tackle all the nonlinearity, and often produces correlated error that requires extra pattern-dependent noise prediction (PDNP) in the branch metric computation of the trellis-based detector \cite{Kavcic_IT, Moon_JSAC}. Secondly, minimizing MSE at the equalizer output does not, in general, guarantee minimum bit error rate (BER) at the detector output. New equalization structure and adaptation criterion are needed for performance improvement.

Different types of neural network equalizers with MSE criterion were investigated in \cite{moon_nn1_1997, rneq, mran}.
In terms of adaptation criterion, maximum equalizer signal-to-distortion ratio (SDR) was proposed in \cite{moon_nn2_1997}, whereas adaptive minimum equalizer BER was employed in \cite{amber}.
However, 
none of these work presented detector output BER evaluation. 
In \cite{osawa2008}, a noise whitening function based on a hybrid genetic algorithm was incorporated in the neural network equalizer design and detector BER was evaluated.
In \cite{nmber}, an equalizer adaptation algorithm was studied that minimizes BER at the Viterbi detector output by considering all relevant bit error sequences when tracing back error paths in the Viterbi detector. 
Both algorithms have a relatively high implementation complexity. 

This paper makes two primary contributions. First, we design nonlinear equalizers (NLE) with multilayer perceptron (MLP) structure. MLP is a class of feedforward neural network, in which adjacent layers act as bipartite graph, and a nonlinear activation function is used in every hidden node of all hidden layers. We demonstrate that NLE show better performance than LE when evaluated on the same criterion as design criterion. In terms of implementation, NLE based on MLP can be readily built upon LE, thanks to the similarity in their structures. Second, we include the soft-output detector in the equalizer design and present a new cost function as adaptation criterion for the equalizer -- cross entropy (CE) between the probability mass function (PMF) of the true bit and the detector's estimate of it. We derive analytically how CE is related to detector BER. Experimental results show that minimizing CE correlates well with minimizing detector output BER, and results in lower detector BER than MSE criterion. Several variations of NLE with cross entropy criterion are studied. The best NLE design gives a 22.76\% detector BER reduction over linear MMSE equalizer. 

This paper is organized as follows. Section~\ref{sec: model} describes the system model and overview. Section~\ref{sec: structure} presents the MLP-based NLE structure. 
Section~\ref{sec: criterion} shows minimizing CE criterion results in ML adaptation. Section~\ref{sec: sims} presents simulation results, and Section~\ref{sec: conclusion} concludes the paper. 

\section{Channel Model and System Overview}
\label{sec: model}

\begin{figure}[t!]
\vspace{-0.1in}
\centerline{\includegraphics[width=\columnwidth]{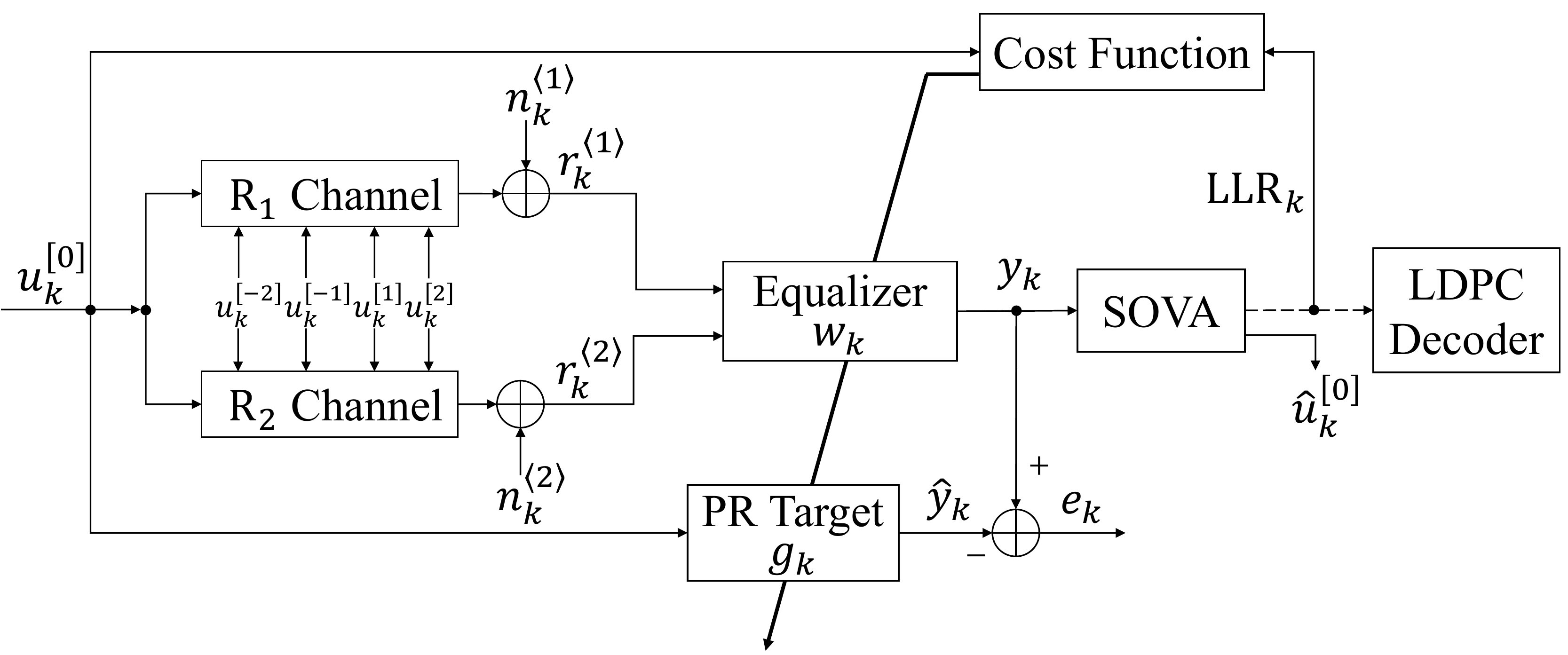}}
\centering
\caption{Discrete-time TDMR system channel model. 
}
\label{fig: channel_model}
\vspace{-0.1in}
\end{figure}

Magnetic recording as a communication channel includes the write and read process. During writing, coded data bits (0 or 1) are recorded on the disk by magnetizing the grains on the storage media in opposite polarities. During reading, the read head scans through each track on the disks and senses the magnetization change.
The recording channel is simulated as a differential system, using the superposition model described in \cite{yang_kurtas}. A readback signal $r_s(t)$ at time $t$ that only captures downtrack ISI effect is simulated as
\begin{equation}
\label{readback}
    r_s(t) = \sum_k b_k \cdot h(t-kT+\Delta t_k)+n(t).
\end{equation}
Here $b_k = (u_k - u_{k-1})/2$, where $u_k$ is the coded bit sequence ($u_k = \pm 1$) and $b_k$ is the transition sequence. $h(t)$ denotes the transition response modeled by the error function $erf(\cdot)$. $n_t$ is electronics noise, assumed to be additive white Gaussian noise (AWGN). The term $\Delta t_k$ represents position jitter, modeled by a Gaussian random variable truncated to $|\Delta t_k| < T/2$, where $T$ is one bit interval. The jitter noise is pattern dependent and colored \cite{kurtas_pdnp}.
The readback signal $r_s(t)$ in \eqref{readback} can be equivalently written as a channel model:
\begin{equation}
    r_s(t) = \sum_k u_k \cdot p(t-kT+\Delta_k)+n(t),
\end{equation}
where $p(t) = h(t)-h(t-T)$ is the channel response, or the dibit response. $p_t$ spans more bits as recording density increases, thus causing severe ISI in high density recording. 

\begin{figure}[t!]
\vspace{-0.1in}
\centerline{\includegraphics[width=.45\columnwidth]{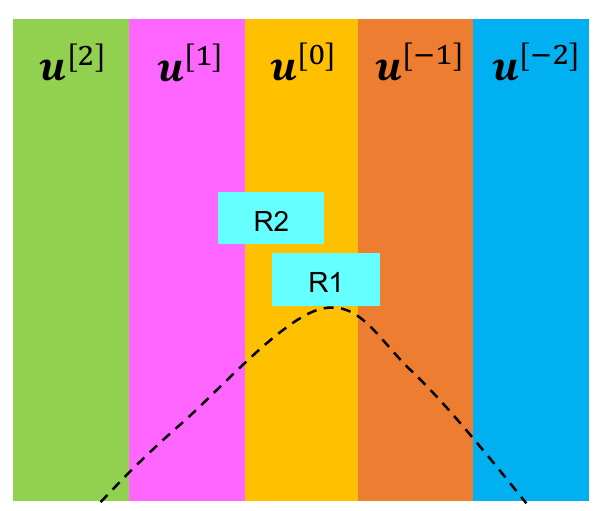}}
\centering
\vspace{-0.03in}
\caption{Reader position on simulated tracks $-2, -1,  \ldots 2$.}
\label{fig: reader}
\vspace{-0.1in}
\end{figure}

The TDMR system employed in this paper is a two-input-one-output system that also mitigates ITI effect from multiple adjacent tracks. Fig.~\ref{fig: channel_model} illustrates the discrete-time TDMR channel model and system overview. 
The system inputs are readback signals from two readers placed with some offset (measured by cross track separation (CTS)) to each other on the same track. As demonstrated in Fig.~\ref{fig: reader}, the readback signal $r^{\langle j\rangle}(t)$ from reader $j, j=1,2$) includes ITI effect from four adjacent tracks:
\begin{equation}
\label{iti}
    r^{\langle j\rangle}(t) = \sum_{i=-2}^{i=2} \lambda^{\langle j\rangle}_i \left[\sum_k u^{[i]}_k \cdot p(t-kT+\Delta t_k) \right]+n^{\langle j\rangle}(t),
\end{equation}
where $\lambda^{\langle j\rangle}_i$ represents the percentage of readback signal captured by reader $j$ from track $i$, and it depends on the reader off-track position. We model cross-track read head response as Gaussian pulse centered at the center of a read head with variance proportional to cross-track density. $\lambda^{\langle j\rangle}_i$ is the area of this Gaussian pulse of reader j that falls under a given track i, and $\sum_i \lambda^{\langle j\rangle}_i = 1$. Track 0 is the center track, and the track of interest. The four tracks centered around track 0 are tracks -2, -1, 1, and 2. We denote the bits ($\pm1$) on track $i$ as $u_k^{[i]}$.

Analog signal $r^{\langle j\rangle}(t)$ is sampled and quantized into digital signal $r^{\langle j\rangle}_k$ by an analog-to-digital converter (ADC). Equivalent discrete-time channel model for $r^{\langle j\rangle}_k$ is:
\begin{equation}
    r^{\langle j\rangle}_k = (\mathbf{p^{\langle j\rangle}}*\mathbf{u})_k + n^{\langle j\rangle}_k,
    j = 1,2,
\end{equation}
where $\mathbf{p^{\langle j\rangle}}$ is the 2D (two dimensional) channel response formed by the dibit response $p(t)$ and $\lambda^{\langle j\rangle}_i$.
$r^{\langle j\rangle}_k$ serves as input to the equalizer. Conventionally, $w_k$ is a linear MMSE equalizer trained to minimize the MSE between the equalized output $y_k$ and a controlled response $\hat{y}_k$, i.e.,
\begin{equation}
    \text{MSE} = E[\mathbf{e}^2] = E[(\mathbf{y}-\mathbf{\hat{y}})^2] = E[(\mathbf{r}*\mathbf{w} - \mathbf{g}*\mathbf{u})^2],
\end{equation}
where $\mathbf{e}$ denotes the error sequence at the equalizer output, and $\mathbf{g}$ is the PR target.
The equalizer output $y_k = (\mathbf{r}*\mathbf{w})_k$ is passed to a trellis-based soft-output detector that is matched to the PR target $\mathbf{g}$. 
The detector employed in this paper is a soft-output Viterbi Algorithm (SOVA) detector \cite{sova}.
The SOVA detector produces both hard estimates $\hat{u}^{[0]}_k$ of the transmitted bits $u^{[0]}_k$ and their corresponding probabilities in the form of log-likelihood ratio (LLR$_k$). Instead of MSE, we propose new cost function for adapting the equalizer $w_k$ -- CE between the probability distribution of the true bit and the SOVA detector's probability estimate of the bit, which can be computed using the detector output LLR. The LLRs are usually passed to a channel decoder, such as Low Density Parity Check (LDPC) decoder, for iterative decoding. LDPC decoding is not considered in this paper.

\section{Equalizer Structure}
\label{sec: structure}

\begin{figure}[t!]
\centering
    \begin{subfigure}[b]{0.45\columnwidth}
         \centering
         \includegraphics[width=\textwidth]{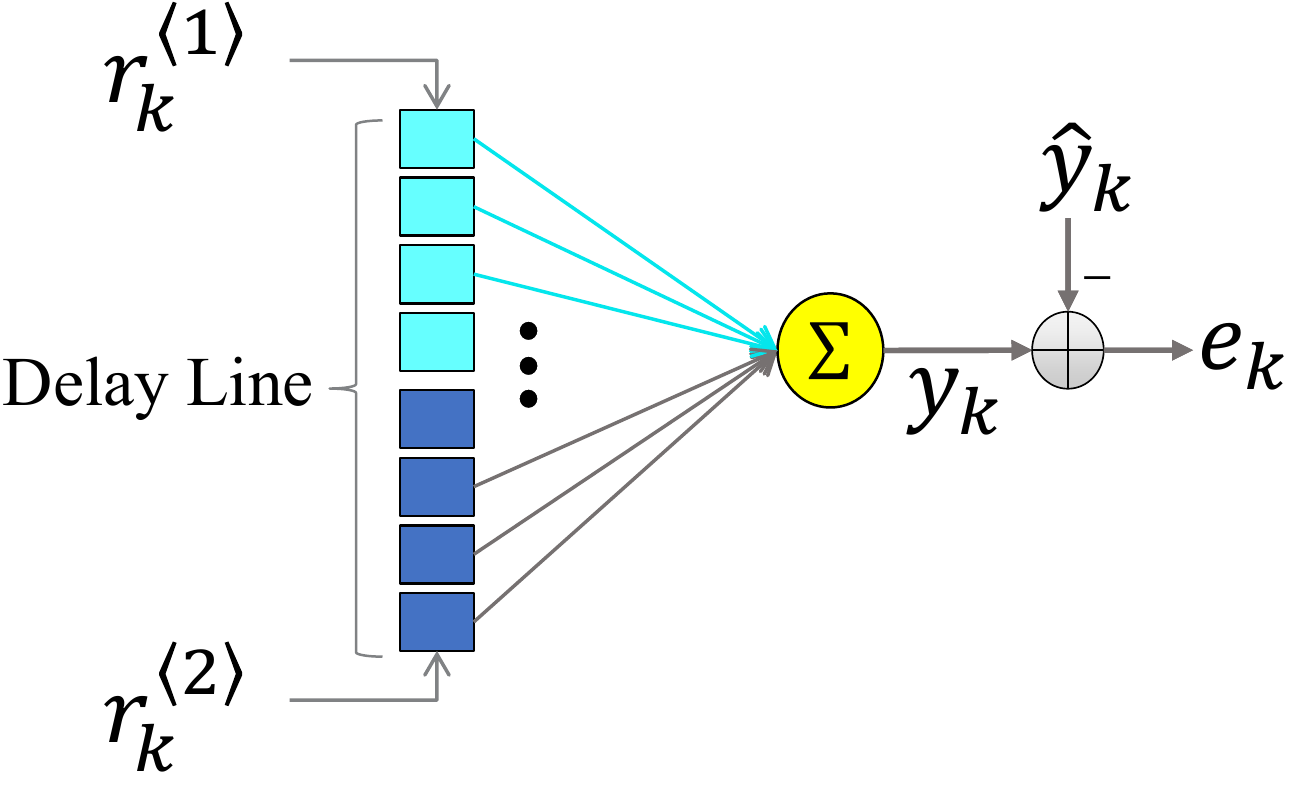}
         \vspace{-0.2in}
         \caption{Linear equalizer.}
         \label{fig: le}
     \end{subfigure}
     \hfill
     \begin{subfigure}[b]{0.53\columnwidth}
         \centering
         \includegraphics[width=\textwidth]{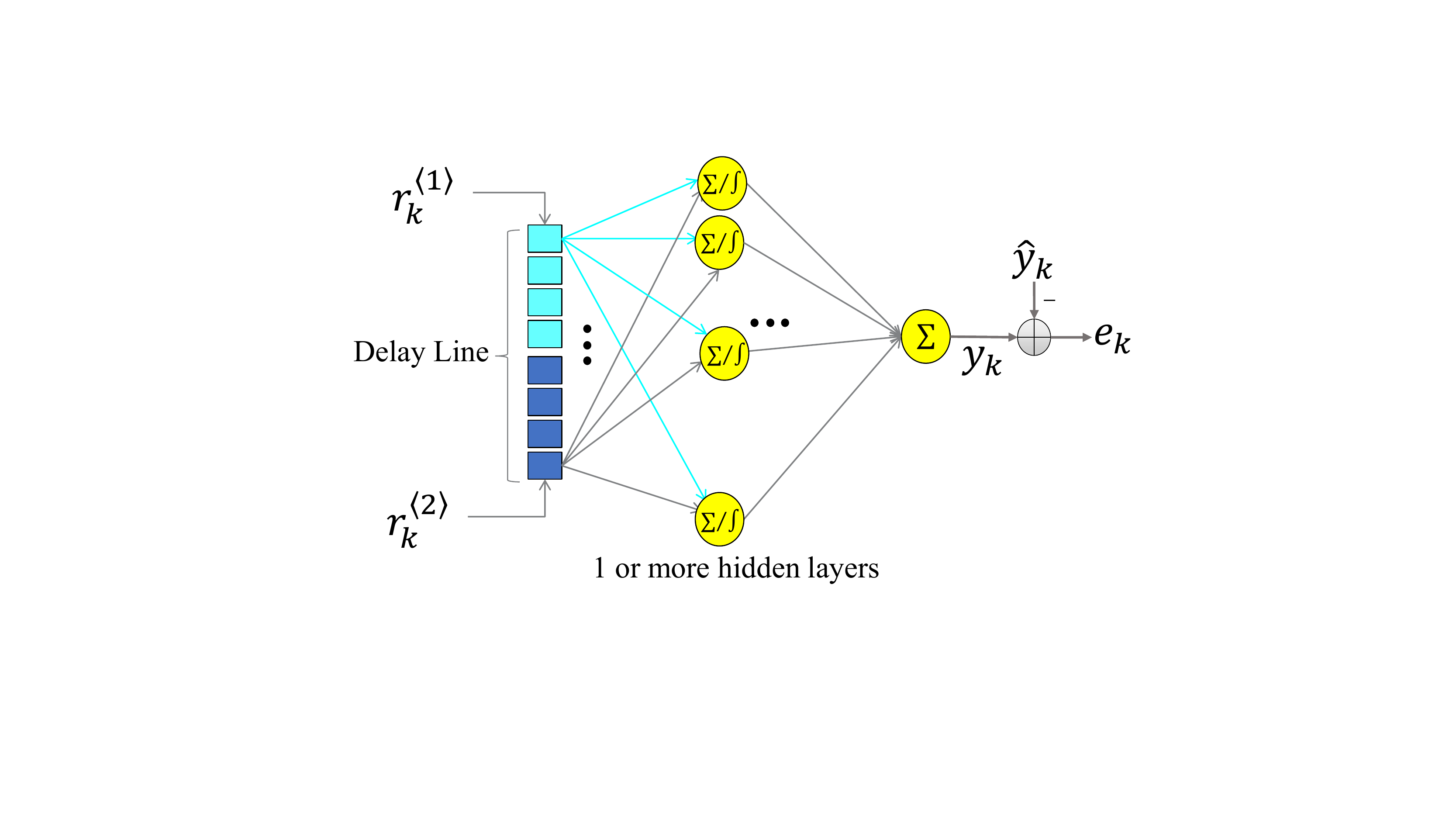}
         \vspace{-0.2in}
         \caption{Nonlinear equalizer.}
         \label{fig: nle}
     \end{subfigure}
     \hfill
\vspace{-0.2in}
\caption{Equalizers as neural networks.}
\label{fig: eqz}
\vspace{-0.1in}
\end{figure}

Conventional equalizers for TDMR are linear FIR filters. A linear FIR equalizer can be illustrated from a neural network point of view. Neural networks consist of layers that are interconnected by weights and biases. In feedforward neural networks, information only flows in the forward direction from input to output. All the nodes in the previous layer are connected to every node in the next layer. Fig.~\ref{fig: le} illustrates the structure of an LE. The delay line indicates the finite memory in the equalizer for handling ISI. The ADC samples $\mathbf{r}^{\langle 1\rangle}$ and $\mathbf{r}^{\langle 2\rangle}$ are fetched as input to the network. Denote the number of input samples from each reader as $D_{in}$. Although only four nodes are shown in Fig.~\ref{fig: le}, $D_{in}$ is typically between 10 and 20. The filtered output $y$ is compared with a reference signal $\hat{y}_k = (\mathbf{u}^{[0]} * \mathbf{g})_k$, and an error signal $e_k$ is computed as $e_k = y_k-\hat{y}_k$. 
Linear filtering is essentially multiplying each of the input nodes by a weight scale and then summing up the results. Thus, filter taps are equivalent to inter-layer weights in the neural network. In other words, we can view a linear FIR equalizer as a vanilla feedforward neural network with no hidden layer and a linear activation function. 

Based on this observation, we propose nonlinear FIR equalizers by extending the above vanilla network to MLP. An MLP is a feedforward neural network that has at least one or more hidden layers with a nonlinear activation function for each of the hidden nodes. Fig.~\ref{fig: nle} shows the structure of a MLP-based nonlinear equalizer. The only difference from the LE in Fig.~\ref{fig: le} is that the NLE in Fig.~\ref{fig: nle} has hidden layer(s) with non-linear activation function. Common activation functions used in the hidden layer include the hyperbolic tangent function $f(x) =tanh(x)$ and rectified linear unit ($ReLU$) $f(x) = \max(0,x)$, etc. 

\begin{figure}[t!]
\vspace{-0.1in}
\centerline{\includegraphics[width=\columnwidth]{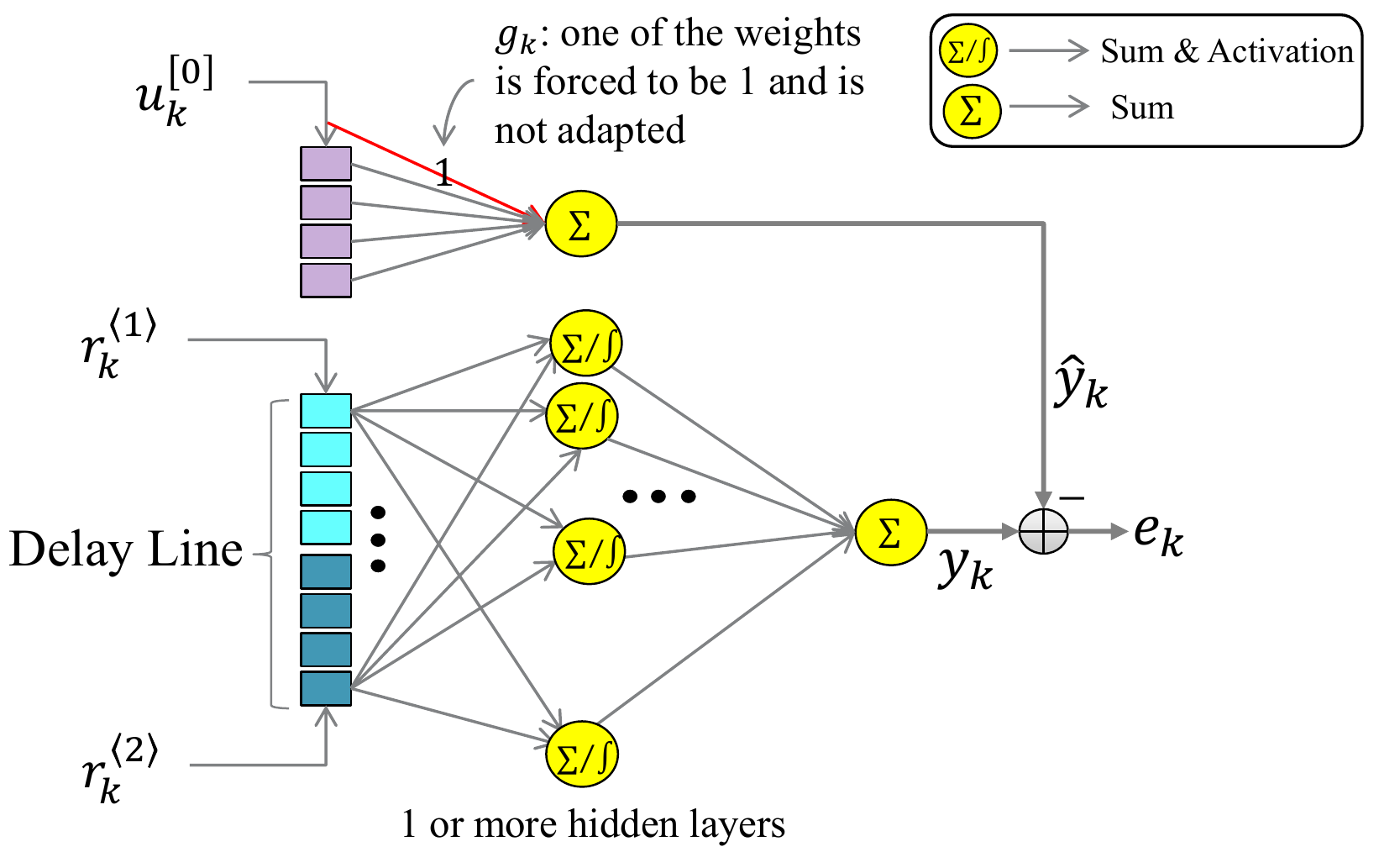}}
\centering
\caption{Nonlinear equalizer with target adaptation.}
\label{fig: ta}
\vspace{-0.1in}
\end{figure}

Because the equalizer aims to shape the received signals into a well-defined response, its performance is influenced by the choice of the PR target $\mathbf{g}$. There are two common options in practical TDMR channels -- either use a fixed PR target, or co-adapt the target with the equalizer taps. Target adaptation (TA) can also be employed in NLE. Fig.~\ref{fig: ta} shows the structure of a NLE with TA. The new network consists of two separate sub-nets. The main sub-net is the MLP for NLE taps and takes ADC samples $\mathbf{r}^{\langle 1\rangle}$ and $\mathbf{r}^{\langle 2\rangle}$ as input. In the other one, input to the sub-net are genie non-return-to-zero (NRZ) bits ($u_k^{[0]}$) and connections represent the PR target taps $g_k$. A 5-tap PR target is considered here. To avoid the situation that all weights are adapted to zero and the equalization error becomes zero, we apply the monic constraint on the PR target, i.e., forcing one of the PR taps to be one. The monic constraint on the equalizer target response is known to be effective in whitening the noise samples at the equalizer output \cite{monic}. The TA sub-net outputs the reference signal $\hat{y}_k$,  whereas the MLP sub-net outputs the equalized signal $y_k$. Their difference is the equalizer error $e_k$.

\section{Adaptation Criterion}
\label{sec: criterion}
Using the MSE criterion, we train an NLE and an LE of comparable size with the same fixed PR target, and compute both equalizer output MSE and detector BER over the same testing data. More details of this experiment are described in Section~\ref{sec: sims}. Results summarized in Table~\ref{tab: mse} proves that minimizing the MSE between equalizer output $y_k$ and controlled response ${\hat{y}_k}$ does not give lowest BER at the detector output. 
We propose to replace MSE with CE as the equalizer adaptation criterion, which is a popular loss function for binary classification problems in deep learning \cite{Goodfellow}.
For bit $u_k^{[0]}$, 
we use the following notation for the true probability $\mathbf{p}_k^{[0]}$ of the bit and detector's estimated probability $\hat{\mathbf{p}}_k^{[0]}$:
\begin{equation}
\begin{split}
    &P_k^- = P(u_k^{[0]} = -1),\;
    P_k^+ = P(u_k^{[0]} = +1),\\
    &\hat{P}_k^- = P(\hat{u}_k^{[0]} = -1),\;
    \hat{P}_k^+ = P(\hat{u}_k^{[0]} = +1),\\
    &\text{Here }P_k^- + P_k^+ = 1,\; \hat{P}_k^-+\hat{P}_k^+=1.
\end{split}
\end{equation}
The cross entropy between $\mathbf{p}_k^{[0]}$ and $\hat{\mathbf{p}}_k^{[0]}$ is defined as: 
\begin{equation}
\label{eqn: ce}
\begin{split}
    &H(\mathbf{p}_k^{[0]},\hat{\mathbf{p}}_k^{[0]})
    = -P_k^- \cdot log(\hat{P}_k^-) 
   -(1-P_k^-) \cdot log(1-\hat{P}_k^-)
\end{split}
\end{equation}
The overall cost function $J$ is the average of CE of all input bits in one forward pass: $J = \frac{1}{K}\sum_{k=1}^K H(\mathbf{p}_k^{[0]},\hat{\mathbf{p}}_k^{[0]})$, where K is the total number of bits in the batch to be estimated.
As shown below, CE is 0 when detector estimates correct decision with high confidence, whereas CE becomes infinity when detector estimates in-correct decision with high confidence.
\begin{equation}
\begin{split}
    H(\mathbf{p}_k^{[0]},\hat{\mathbf{p}}_k^{[0]}) = 0, \text{ when }
    &\begin{cases}
    u_k^{[0]} = -1,\: \hat{P}_k^- = 1\\ 
    u_k^{[0]} = +1,\: \hat{P}_k^+ = 1
    \end{cases} \\
    H(\mathbf{p}_k^{[0]},\hat{\mathbf{p}}_k^{[0]}) = \infty, \text{ when }
    &\begin{cases}
    u_k^{[0]} = -1,\: \hat{P}_k^- = 0 \\ 
    u_k^{[0]} = +1,\: \hat{P}_k^+ = 0,
    \end{cases}
\end{split}
\label{eqn: extreme}
\end{equation}
Therefore minimizing CE can be used as an adaptation criterion for adapting any system parameter and does truly reflect the quality of detected bits.

In most classification problems that uses CE loss, the neural network estimates $\hat{\mathbf{p}}_k^{[0]}$. We propose to estimate $\hat{\mathbf{p}}_k^{[0]}$ using standard detector, by means of detector LLR.
Let's see what effect minimizing CE has on detector LLR estimates. We first represent CE in ~\eqref{eqn: ce} in terms of LLR.
\begin{equation}
\begin{split}
    &\text{LLR}_k = \log(\frac{\hat{P}_k^+}{\hat{P}_k^-}),\;
    \hat{P}_k^- = \frac{1}{1+e^{\text{LLR}_k}},\;
    \hat{P}_k^+ = \frac{e^{\text{LLR}_k}}{1+e^{\text{LLR}_k}},\\
    &H(\mathbf{p}_k^{[0]},\hat{\mathbf{p}}_k^{[0]})
    = -P_k^- \cdot \log(\frac{1}{1+e^{\text{LLR}_k}})
    - P_k^+ \cdot \log(\frac{e^{\text{LLR}_k}}{1+e^{\text{LLR}_k}})
\end{split}
\end{equation}
With CE represented in terms of LLR, we can adapt LLR using criterion of minimizing CE. To do this, we find the gradient of CE w.r.t. LLR and adapt LLR via steepest decent algorithm. We can show that 
\begin{equation}
\begin{split}
    \frac{\partial(H(\mathbf{p}_k^{[0]},\hat{\mathbf{p}}_k^{[0]}))}{\partial({\text{LLR}_k})} 
    &= 
    P_k^- \cdot \frac{e^{\text{LLR}_k}}{1+e^{\text{LLR}_k}} -  
    P_k^+ \cdot \frac{1}{1+e^{\text{LLR}_k}} \\
    &= 
    \begin{cases}
    \hat{P}_k^+, \text{ when } u_k^{[0]} = -1 \\
    -\hat{P}_k^-, \text{ when } u_k^{[0]} = +1
    \end{cases}
\end{split}
\label{eqn: ce2}
\end{equation}
The equation for updating LLR from step $t$ to step $t+1$ using steepest gradient descent is then
\begin{equation}
\begin{split}
    {\text{LLR}_k}(t+1) &= {\text{LLR}_k}(t) - \mu \cdot \hat{P}_k^+, \text{ when } u_k^{[0]} = -1 \\
    {\text{LLR}_k}(t+1) &= {\text{LLR}_k}(t) + \mu \cdot \hat{P}_k^-, \text{ when } u_k^{[0]} = +1,
\end{split}
\label{eqn: derivative}
\end{equation}
where $\mu$ is the learning rate.
\eqref{eqn: derivative} shows the impact of the CE loss when using gradient descent for adaptation. When the true bit $u_k^{[0]} =-1$, minimizing CE at step $t$ lowers $\text{LLR}_k(t)$ by an amount proportional to $\hat{P}_k^+$, the detector's probability for incorrect decision $\hat{u}_k^{[0]} = +1$. 
Similarly, when the true bit $u_k^{[0]} =+1$, minimizing CE at step t increases $\text{LLR}_k(t)$ by an amount proportional to $P_k^-$, the detector's probability for incorrect decision $\hat{u}_k^{[0]} = -1$.  
Eventually, when CE is minimized, 
LLR$_k$ for true bit $u_k^{[0]}  = -1$ becomes more negative and LLR$_k$ for true bit $u_k^{[0]}  = +1$ become more positive. 
Hence, the proposed adaptation based on minimizing CE is equivalent to maximizing the likelihood of detected bits (ML adaptation).

Next we show how CE is related to the equalizer taps. The detector employed in this paper is a SOVA detector \cite{sova}, a modified version of the Viterbi Algorithm that outputs LLR. The Viterbi Algorithm is a trellis-based ML sequence detector that is matched to the PR target $\mathbf{g}$, and has number of trellis states equal to $2^{length(g)-1}$. At stage $k$ of the trellis, the branch metric (BM) $\text{BM}_k^i$ for branch $i$ is calculated as the squared Euclidean distance between the equalized output $y_k$ and controlled response $\hat{y}_k^i$ corresponding to branch $i$, i.e., $\text{BM}_k^i = (y_k-\hat{y}_k^i)^2$, where $\hat{y}_k^i = (\mathbf{g}*\mathbf{u}^i)_k$ and $\mathbf{u}^i$ denotes the assumed input bits on branch $i$. To decide the surviving path that arrives at each state, it performs three operations: add, compare and select (ACS) \cite{va}. At any stage, the path metric (PM) for each state is the sum of the BM of all the branches on the survivor path leading to that state. In our implementation, the SOVA detector traces back error paths and approximates LLR using the MAX-Log-MAP algorithm \cite{maxlogmap}:
\begin{equation}
    \text{LLR}_k = \text{LLR}(u_k^{[0]}) = \min_{\substack{\text{all paths} \\ \ni u_k^{[0]}=+1}}(\text{PM}) - \min_{\substack{\text{all paths} \\ \ni u_k^{[0]}=-1}}(\text{PM})
\end{equation}
In other words, the LLR of bit $u_k^{[0]}$ is decided by the path metric difference (PMD) between the most likely (ML) path that has $u_k^{[0]} = -1$ and the most likely (ML) path that has $u_k^{[0]} = 1$. Hence, the relationship between CE and the equalizer taps (and PR target taps when TA is used) can be shown in the following computation graph.
\begin{equation}
\begin{rcases}
\text{FIR taps}\\
\text{Target taps}
\end{rcases}
\longleftarrow \text{BM}
\longleftarrow \text{PM}
\longleftarrow \text{PMD}
\longleftarrow \text{LLR}
\longleftarrow \text{CE}
\end{equation}
Backpropagation\cite{bp} is used in the gradient descent optimization algorithm to effectively train the neural network in the equalizer, i.e., 
\begin{equation}
    \frac{\partial(\text{CE})}{\partial(\text{tap})} = \frac{\partial(\text{CE})}{\partial(\text{LLR})} \cdot 
    \frac{\partial(\text{LLR})}{\partial(\text{PMD})} \cdot 
    \frac{\partial(\text{PMD})}{\partial(\text{PM})} \cdot 
    \frac{\partial(\text{PM})}{\partial(\text{BM})} 
    \cdot 
    \frac{\partial(\text{BM})}{\partial(\text{tap})},
\end{equation}
where tap is an FIR tap (or PR tap when TA is used). Existing deep learning software such as PyTorch and TensorFlow have built-in packages that performs automatic differentiation and backpropagation.
We point out that the above analysis is not restricted to employing the SOVA detector. The Bahl-Cocke-Jelinek-Raviv (BCJR) detector, a symbol-by-symbol maximum {\em a posteriori} (MAP) detector is an alternative detector for use in TDMR systems \cite{tdmr}. When BCJR is used in place of SOVA, LLR magnitudes can again be approximated using MAX-log-MAP.
The computation graph and back propagation still hold, except that the PMD and BM are replaced by $\alpha$, $\beta$, and $\gamma$ as defined in \cite{bcjr}.

\section{Simulation Results}
\label{sec: sims}

\begin{table}[t!]
\centering
\caption{NLE v.s. LE using MSE criterion}
\label{tab: mse}
\begin{tabular}{|c|c|c|c|}
\hline
\textbf{\begin{tabular}[c]{@{}c@{}}Equalizer \\ Structure\end{tabular}} & \textbf{\begin{tabular}[c]{@{}c@{}}Adaptation \\ Criterion\end{tabular}} & \textbf{MSE} & \textbf{Detector BER}  \\ \hline
LE & MSE & 7.54 & 0.0180 \\ \hline
NLE & MSE & 5.83 & 0.0297 \\ \hline
\end{tabular}
\vspace{-0.1in}
\end{table}

\begin{figure}[t!]
\centerline{\includegraphics[width=0.9\columnwidth]{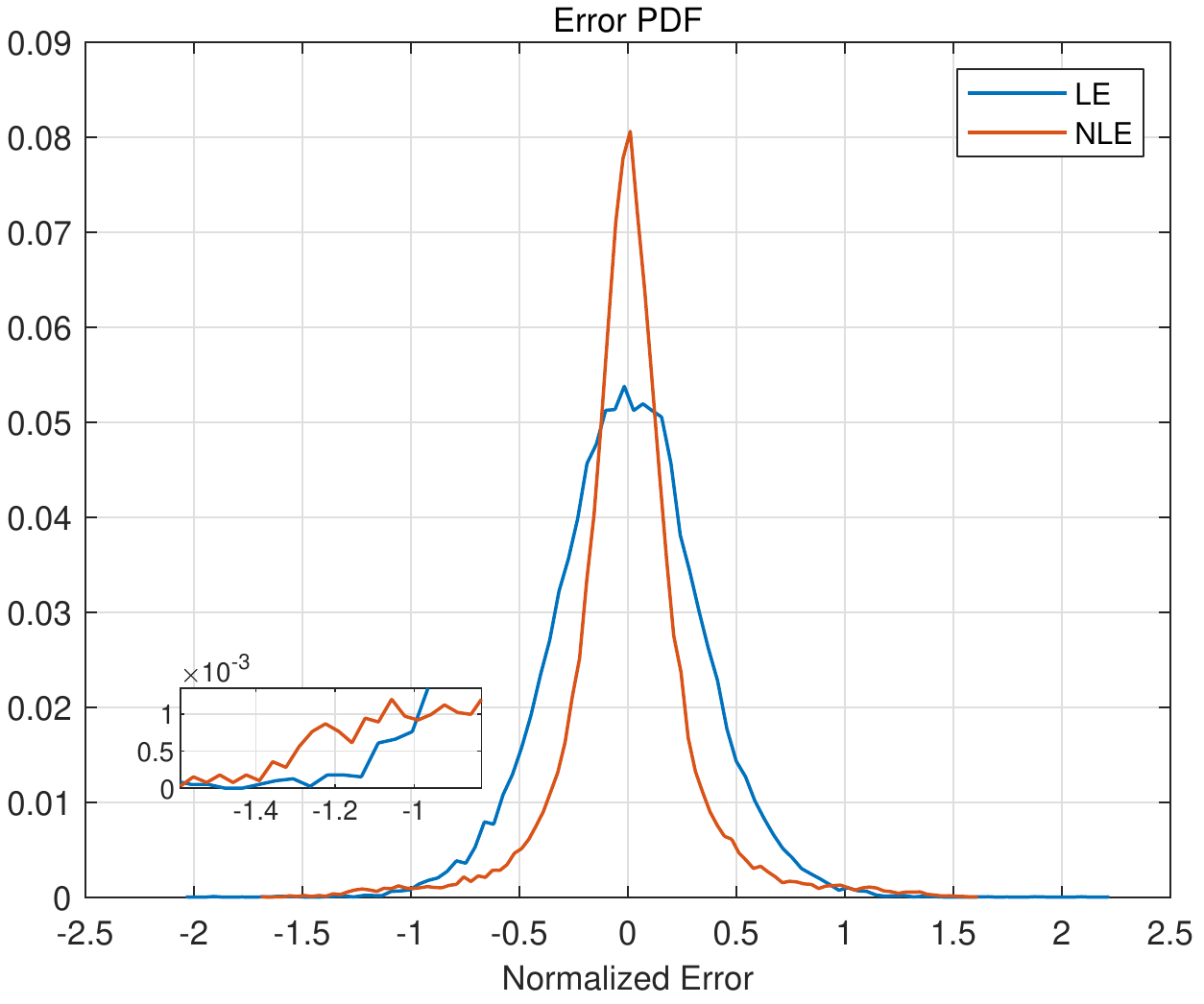}}
\centering
\vspace{-0.1in}
\caption{Equalizer output error PDF.}
\label{fig: pdf}
\vspace{-0.2in}
\end{figure}

\begin{figure*}[t!]
\centering
    \begin{subfigure}[b]{0.32\textwidth}
         \centering
         \includegraphics[width=0.9\textwidth]{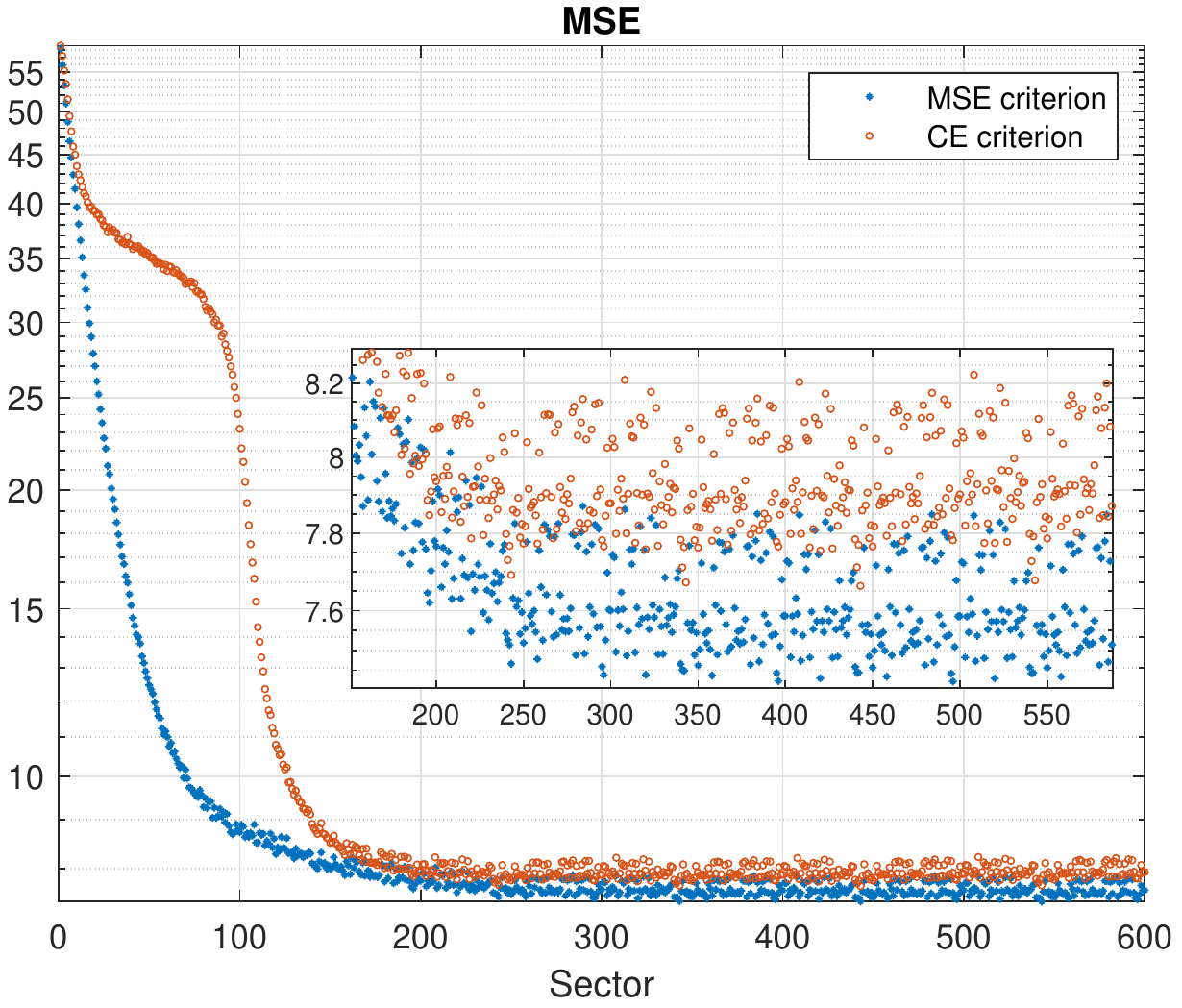}
         \vspace{-0.1in}
         \caption{}
         \label{fig: mse}
     \end{subfigure}
     \hfill
     \begin{subfigure}[b]{0.32\textwidth}
         \centering
         \includegraphics[width=0.9\textwidth]{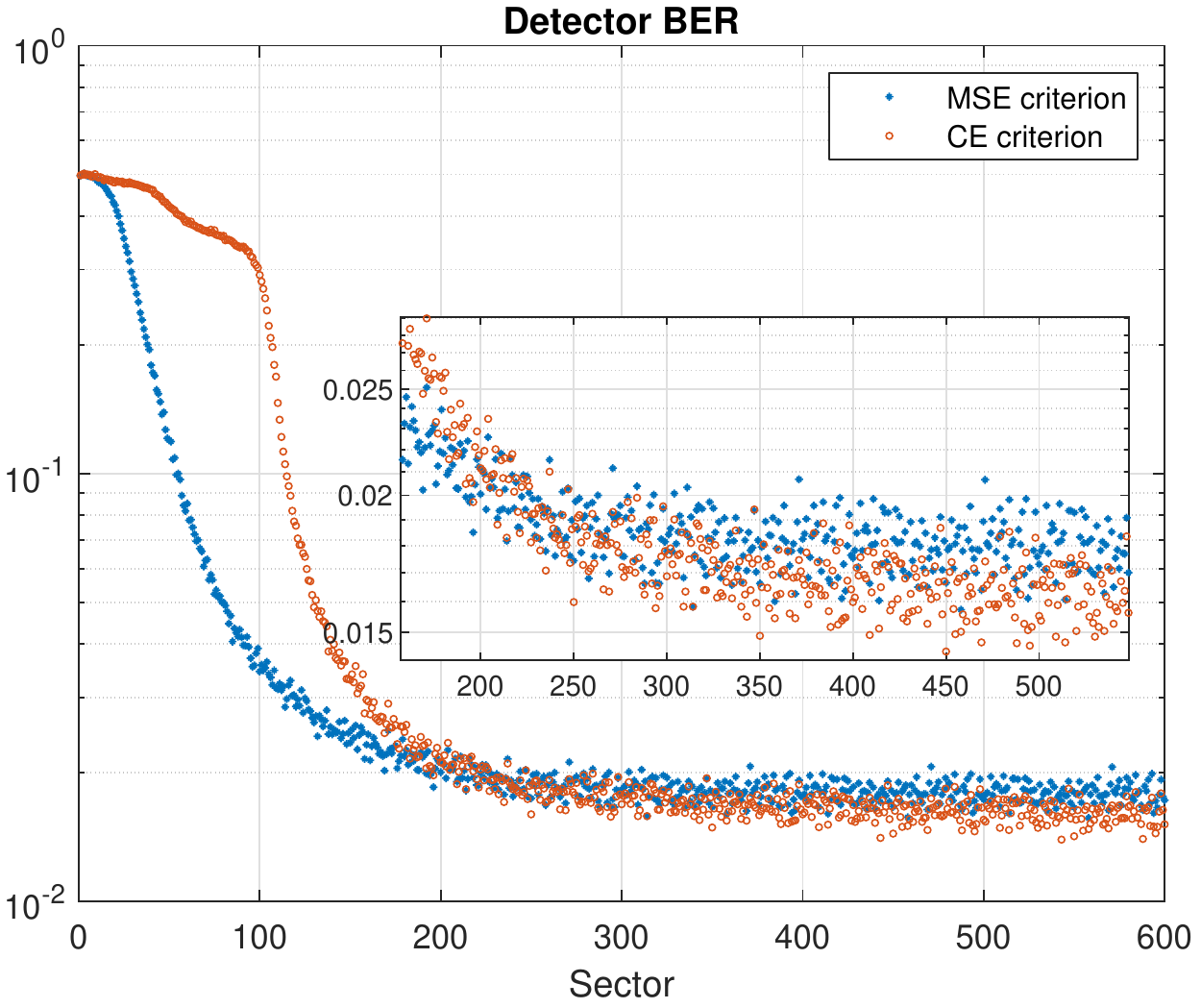}
         \vspace{-0.1in}
         \caption{}
         \label{fig: ber}
     \end{subfigure}
     \hfill
     \begin{subfigure}[b]{0.32\textwidth}
         \centering
         \includegraphics[width=0.9\textwidth]{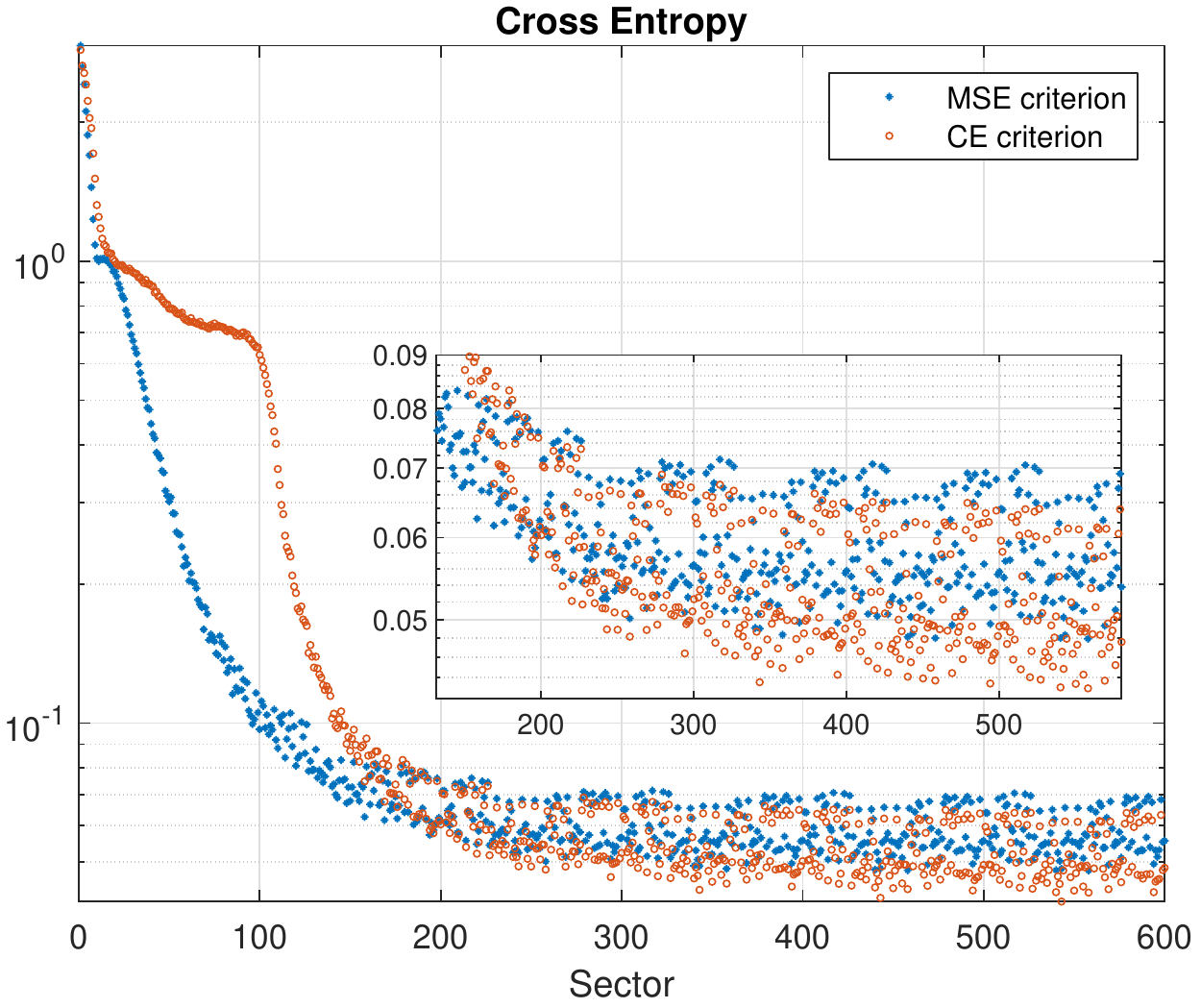}
         \vspace{-0.1in}
         \caption{}
         \label{fig: ce}
     \end{subfigure}
     \hfill     
\vspace{-0.1in}
\caption{Adaptation curve for LE with CE criterion. a) MSE, b) BER, c) CE.}
\label{fig: ce_le}
\vspace{-0.2in}
\end{figure*}

The proposed two-reader TDMR detection system is evaluated on simulated HDD data with 30\% CTS between the two readers positioned symmetrically around track 0. The values of $\lambda^{\langle j\rangle}_i, i = -2, \ldots, 2$ in \eqref{iti} are $[3.54e-05, 0.1514, 0.8207, 0.0279, 5.96e-07]$ for reader 1, and same values in reverse order for reader 2 (refer to Fig.~\ref{fig: reader} for illustration). 
The raw BERs of ADC samples $r^{\langle 1\rangle}_k$ and $r^{\langle 2\rangle}_k$ are roughly 11\%. The system is implemented in open source library PyTorch. We train the MLP using the Adam optimizer, a commonly used adaptive learning rate optimization algorithm. We use default values for $\beta_1, \beta_2, \epsilon$ as defined in \cite{kingma2014adam}, and choose learning rate $\mu = 10^{-3}$. Gradient descent is performed every mini batch consisting of 1,024 input examples. 

A total of 100 sectors of ADC samples $r^{\langle 1\rangle}_k$ and $r^{\langle 2\rangle}_k$ are available. Each sector contains $N_s = 39,512$ bits, and same number of $r^{\langle 1\rangle}_k$ and $r^{\langle 2\rangle}_k$.
To accelerate the training, the ADC samples are normalized per sector and per reader so that they have zero mean and unit standard deviation. The equalizer architecture is denoted as $[2*D_{in},H_1,...,H_l, 1]$, where the number of input ADC samples from each reader $D_{in} = 11$, $H_k$ is the number of hidden nodes in the $k^{th}$ hidden layer, and 1 is the number of output node. Our choice of $l$ is up to two. We form the neural network input using a sliding window fashion down the track, and the total number of input examples are $100(N_s-D_{in}+1)$.

In experiments using MSE criterion, we find using 1 sector for training is adequate. For CE criterion, we initially use 2 sectors for training and the remaining 98 sectors for inference, and find no significant performance difference between them. This indicates no over-fitting has occurred, which is expected given that our neural network has relatively small number of learnable parameters (the largest neural network we designed has a total of 171 learnables), compared to the number of bits in a sector. For better performance, all 100 sectors are used for every epoch in results  presented in this section using CE criterion. At the beginning of the next epoch, the same 100 sectors are cycled through repetitively. The BER values given in this section are final results that no longer decrease with more training epochs. The learnables are randomly initialized in all experiments.

\textbf{Equalizer Structure:} We first compare LE and NLE structure, using MSE criterion. The NLE has structure [22-4-1] (total 97 parameters) and uses the $tanh$ activation function. The LE size is chosen as [98,1] so that they have comparable complexity, although we find no significant performance difference between structures [98,1] and [22,1]. LE with MSE criterion is the linear MMSE equalizer. Both systems use a fixed PR target of [4,7,1], whose D-transform is $G(D) = 4+7D+D^2$. 
The system is trained on sector 1, and tested on sectors 2 through 6. MSE and BER results are averaged over all six sectors.
Table~\ref{tab: mse} shows that the NLE output has a lower MSE than LE output, but results in a higher detector BER.
The reason is revealed in Fig.~\ref{fig: pdf}, which shows that the probability density function (PDF) of NLE output errors (approximated by error histogram) is more concentrated around 0, and consequently leads to lower MSE. However, the area under the tail of NLE error PDF is larger than that of LE (shown in the zoomed-in plot in the same figure). Because the bits in the tail suffer from larger magnitude equalization error, the detector tends to make more wrong decisions on these bits rather than those bits centered around 0 error in the PDF. 
The above results show that 1) NLE outperforms LE under the same adaptation criterion, and 2) a new criterion is needed that is coherent with detector BER.

\textbf{Adaptation Criterion:} We next compare CE and MSE criterion on the same equalizer structure. Each criterion is used to train an LE of size [22,1], and we plot for each sector the values of MSE, detector BER, and CE during adaptation. Sectors 1 through 100 are cycled through the training. The sector number $100*m_e+m_s$ in the horizontal axis of all three plots means sector $m_s$ in the $(m_e+1)^{th}$ epoch. In Fig.~\ref{fig: ce_le}, CE criterion results in lower final CE (Fig.~\ref{fig: ce}) and BER (Fig.~\ref{fig: ber}) than MSE criterion, but a higher final MSE (Fig.~\ref{fig: mse}). The CE curve is consistent with the BER curve between the two criteria, whereas the MSE curve is not. The final detector BERs from MSE and CE criterion are 0.0179 and 0.0159. The conclusion is that CE correlates well with detector BER, and is a better adaptation criterion than MSE. Using the CE criterion, we compare equalizer structure again and find that BERs from LE and NLE are 0.0159 and 0.0128. This shows NLE is still superior to LE under CE criterion.

We now present results several variations of NLE with CE criterion. Although adaptation curves are not shown, they exhibit similar behaviour to Fig.~\ref{fig: ce_le}.

\begin{table}[t!]
\caption{Detector BER from different NLE with CE}
\vspace{-0.1in}
\label{tab: nle}
\begin{center}
\centering
\begin{tabular}{|l|l|l|l|}
\hline
\textbf{NLE Structure} & \textbf{activation} & \textbf{TA} & \textbf{BER} \\ \hline
[22-6-1] & tanh & No & 0.0128 \\ \hline
[22-6-1] & tanh & Yes & \textbf{0.0112} \\ \hline
[22-6-1] & ReLU & Yes & 0.0121 \\ \hline
[22-4-1] & tanh & Yes & 0.0123 \\ \hline
[22-8-1] & tanh & Yes & 0.0117 \\ \hline
[22-6-3-1] & tanh & Yes & 0.0123 \\ \hline
[22-6-4-1] & tanh & Yes & 0.0139 \\ \hline
\end{tabular}
\end{center}
\vspace{-0.2in}
\end{table}

\begin{table}[t!]
\centering
\caption{Overall detector BER comparison}
\label{tab: best}
\begin{tabular}{|c|c|c|}
\hline
\textbf{\begin{tabular}[c]{@{}c@{}}Equalizer \\ Structure\end{tabular}} & \textbf{\begin{tabular}[c]{@{}c@{}}Adaptation \\ Criterion\end{tabular}} & \textbf{Detector BER} \\ \hline
LE & MSE & 0.0145 \\ \hline
LE & CE & 0.0137 \\ \hline
NLE & CE & 0.0112 \\ \hline
\end{tabular}
\vspace{-0.2in}
\end{table}

\textbf{Target Adaptation:} TA is compared against the fixed PR target [4,7,1] on NLE structure [22-6-1] with $tanh$ activation. For TA, we consider a 5-tap monic PR target [1,x,x,x,x], where the monic tap 1 is imposed on current bit.
The first two rows in Table~\ref{tab: nle} show that TA outperforms fixed target. The final adapted target is [1, 0.5367, 0.0781, -0.1535, 0.0347]. 
The final detector BER is 0.0112. This is the lowest BER achieved among all variations of NLE with CE criterion, shown in the third row of Table~\ref{tab: best}. 

\textbf{Other variations:} Results on other variations of NLE with TA using CE criterion are shown in Table~\ref{tab: nle}. Rows 2 and 3 indicate that the $tanh$ activation outperforms $ReLU$, on [22-6-1] NLE. Rows 2,4 and 5 suggest that NLE with 6 hidden nodes ([22-6-1]) wins over [22-4-1] and [22-8-1], with $tanh$ activation. Comparing rows 2, 6 and 7, it appears one more hidden layer does not yield better performance. This could mean that NLE with more than one hidden layer is a overly complex equalization model for magnetic recording channels.
 
\textbf{Best Result:} We summarize the final detector BER values from three combinations of equalizer structure and adaptation criterion in Table~\ref{tab: best}. Both the LE with MSE and CE criterion has structure [22, 1] with TA. We find experimentally that using $D_{in}>22$ in LE does not give significant performance improvement. The NLE with CE criterion refers to the best NLE structure -- [22-6-1] with $tanh$ activation and TA. 
Detector BER converges after 2 epochs for LE with MSE, 14 epochs for LE with CE, and 17 epochs for NLE with CE.
LE with CE criterion achieves 5.52\% lower detector BER than LE with MSE criterion, and the best NLE with CE criterion gives further 18.25\% BER reduction compared to LE with CE criterion. Overall, the best NLE with CE criterion yields 22.76\% lower detector BER than the linear MMSE equalizer.


\section{Conclusion}
\label{sec: conclusion}
We design MLP-based NLE for TDMR channels and propose to adapt it with CE between the true probability of the bit and detector’s estimate of it. We show that NLE is a better structure than LE, and CE is a superior criterion to MSE, in terms of detector BER performance. Further BER reduction is anticipated from hyperparameter fine tuning, longer training epochs and branch metric modification and co-adapting in the Viterbi detector.

\bibliographystyle{IEEEtran}
\bibliography{ref}

\end{document}